# Scanning the IPv6 Internet: Towards a Comprehensive Hitlist

Oliver Gasser, Quirin Scheitle, Sebastian Gebhard, Georg Carle
Technische Universität München
Chair of Network Architectures and Services
Email: {gasser,scheitle,sgebhard,carle}@net.in.tum.de

*** *Extended Version* ***

***Abstract*—Active network measurements constitute an important part in gaining a better understanding of the Internet. Although IPv4-wide scans are now easily possible, random active probing is infeasible in the IPv6 Internet. Therefore, we propose a hybrid approach to generate a hitlist of IPv6 addresses for scanning: First, we extract IPv6 addresses from passive flow data. Second, we leverage publicly available resources such as rDNS data to gather further IPv6 addresses. Third, we conduct traceroute measurements from several vantage points to obtain additional addresses. We perform multiple active measurements on gathered IPv6 addresses and evaluate response rates over time. We extensively compare all IPv6 address sources. In total we found 150M unique IPv6 addresses over the course of four weeks. Our hitlist covers 72% of announced prefixes and 84% of Autonomous Systems. Finally, we give concrete recommendations to maximize source efficiency for different scan types.**

## I. Introduction

The deployment of IPv6 is quickly gaining momentum: By the end of 2015 native IPv6 traffic to Google surpassed 10% [12]. To better understand the current state of the Internet, active measurements are an important tool. Since the address space in IPv6 is orders of magnitude larger than in IPv4 a complete scan is not feasible. Therefore smart address selection techniques are needed to probe for potentially active IPv6 addresses. By gathering addresses from both active and passive measurements, this paper presents a first step into systematic selection of IPv6 addresses for scanning. Over the course of four weeks we gathered flow data from a major uplink of universities and scientific institutions and sampled flow data from a large European IXP. During this collection process we conducted multiple active measurements to all observed IPs. This allowed us to evaluate the passive flow data with regards to the response rate, in addition to AS and prefix coverage. As active sources we selected various DNS data sets [22], [23], [24], the Alexa Top 1 Million list [2], and the CAIDA DNS Names data set [5]. We carried out traceroutes to all IPs collected from active sources in order to find additional IPs pertaining to routers. Finally, we give concrete advice on the usefulness of sources depending on scan type.

**Contributions.** This paper makes the following contributions: (a) We extensively compare different sources for IPv6 addresses. (b) We research time stability of addresses depending on source, observed protocol type, and scan type. (c) We give specific advice for efficiently combining sources depending on the intended scan type.

Highlights of our results include: (a) Uplink vantage points, in comparison to IXPs, see much less unique IPs, more tunneling techniques and much more DNS traffic. (b) ICMPv6 response rate is in almost all cases higher than in-protocol response rates. This even applies for the Alexa 1M list, which should offer a 100% response rate to HTTP queries. This rule does not apply to IPs observed on peer-to-peer ports (e.g. udp49001), likely due to home routers dropping `ICMPv6 echo request` packets. (c) Given the quick decline in responsiveness and dominance of privacy extensions, the number of observed IPs is a pointless metric. Building on [21], we suggest to focus on IPs observed as stable for at least a week. (d) Passive sources provide great coverage but are difficult to obtain and operate. As only 15% of ASes and 31% of /64 prefixes were found exclusively in passive sources, they might be omitted where completeness is not critical.

**Outline.** In Section II we discuss related work and compare it to our approach. Section III explains our method, followed by a list of data sources and the implementation description in Section IV. In Section V we evaluate the gathered hit lists and synthesize those insights into specific recommendations per scan type, concluding in Section VI.

## II. Related Work

In this section we first survey work on IPv6 deployment. Then we present studies which focus on hitlist generation. Finally, we review work on Internet scanning.

**IPv6 Deployment and Classification.** Malone in 2008 [19] classified the Interface IDs of IPv6 addresses from server logs and active traceroutes. He reports a majority of EUI-64 or manually configured ("low") addresses for server logs and ≈90% "low" addresses for traceroute respondents. Czyz et al. in 2014 [6] globally measured IPv6 adoption, leveraging 10 global data sets such as Alexa 1M DNS lookups and passive traffic samples. They report a traffic profile of over 95% of IPv6 traffic being user traffic (HTTP or HTTPS). In 2015, Plonka and Berger [21] released their work on IPv6 address classification based on Akamai server logs. Over the course of one year, they recorded clients from 133 countries and over 4,000 Autonomous Systems, which they classified for temporal stability and spatial density. They find only $\approx 4\%$ of addresses to be stable for 4 days or more. They also observe about $\approx 1\%$ EUI-64 addresses. Of the unstable EUI-64 addresses about $62\%$ roamed between /64 prefixes. CAIDA's Archipelago project randomly probes all routed /48 or shorter prefixes in a distributed and time-spread manner from all its monitors. We use their dataset [5] both as an input and for evaluation purposes.

**Hitlist Generation.** Although some IPv6 hitlist generation methods are publicly known (e.g., RFC 7707), this is to the best of our knowledge the first work evaluating IPv6 hitlist generation. However, there is a long history of work on IPv4 hitlist generation: In 2000, Govindan et al. [13] published their work on using random informed probing to self-generate a hitlist aimed on Internet structure discovery. In 2001, Huffaker et al. [17] started using first hitlists generated from web server logs and traffic traces, involving extensive manual processing. In 2008, Sherwood et al. [26] presented a tool for probing a fixed address per subnet in a distributed manner. Permitted by advances in available data rate and scanning tool performance, Fan et al. [9] in 2010 published their work on creating a complete and stable IPv4 hitlist. For this, they repeatedly scanned the entire IPv4 address space and singled out IP addresses that were stable in responsiveness. They share this list, but warn that this might introduce both systematic bias and potential endpoint overload by several researchers probing the same set of IP addresses repeatedly.

Summarizing, IPv4 Internet probing moved from specific random or informed probing into a then sparsely populated Internet to repeatedly fully probing the now densely populated IPv4 space and recording time-stable hosts for efficient hitlists. We believe that the IPv6 Internet will stay sparsely populated for years to come, and hence suggest informed repeated probing to generate a core set of stable and active addresses.

**Internet scans.** Internet scans have long been used by researchers to derive a better understanding of network protocols. In recent years these types of scans have especially been utilized to survey the security of network protocols. In 2011 Holz et al. [16] studied the SSL landscape using active and passive measurements over a period of over 1.5 years. They found that the SSL PKI was in a sorry state, e.g. due to incorrect certification chains or invalid certification names. In 2012 Heninger et al. [15] evaluated the cryptographic security of TLS and SSH. They performed large-scale measurements to gather TLS certificates and SSH host keys. The authors investigated the issue of malfunctioning random number generators resulting in predictable RSA and DSA keys. Durumeric et al. [7] in 2013 published a long-term study of the SSL PKI. They performed 110 Internet-wide active measurements over a period of 14 months and found practices that put the certification system at risk. In 2014 Gasser et al. [10] conducted their own Internet-wide scans and confirmed many of Heninger et al.'s findings for SSH. Furthermore, they evaluated the phenomenon of duplicate yet strong keys.

## III. Approach

In this section we describe our approach and give an insight in the ethical considerations that we followed during the measurements. Generally, we first gather IP addresses from various sources. We then actively probe those addresses for responsiveness of different protocols over time. From traceroute probes, we gain additional router addresses. Finally, we evaluate the usefulness of all sources for different scan types.

**Sources.** For this work, we gathered 3 types of sources:

**Passive:** Collecting flow data of IPv6 Internet traffic.

**Active:** Obtaining static files and performing additional steps, such as DNS resolution.

**Traceroute:** Conducting traceroutes, yielding router IPs.

**Filtering.** To probe and evaluate our hitlists, we first filter out undesired addresses: First, we drop flows generated by our co-located measurement machine. This avoids introducing a bias caused by our own measurements. Second, we separate the flows into individual addresses and remove duplicates. Third, we filter the addresses against the daily updated fullbogon list of TeamCymru[1]. Fourth, we filter out IANA special reserved IP ranges[2]. Fifth, we filter out scans into our own networks. Sixth, we whitelist the remainder against the current CAIDA's Prefix to AS (pfx2as) list [4] to only scan announced prefixes. Seventh, we whitelist routes actually announced in BGP. Finally, we remove IP addresses from blacklisted networks.

**IPv6 Scanning Tool.** To actively probe large amounts of IPv6 addresses, we need a tool capable of high-volume scans at different protocols. Tools such as `ping` or `nmap` are capable of probing IPv6 hosts, but lack performance. Recently, tools such as `masscan` [14], and `zmap` [8] made it possible to scan the complete IPv4 Internet in less than five minutes [1]. Since these tools lack support for IPv6 network measurements we decided to extend zmap to make it IPv6-capable. Specifically, we added generic IPv6 support to zmap and furthermore ported some probe modules to IPv6. The modified version of zmap allows us to scan hosts for ICMPv6 (Echo Request), IPv6 TCP SYN (any port) and IPv6 UDP (any port and payload). The modified version is available on our website [11].

**Ethical Considerations.** Active network measurements can be interpreted as an attack, resulting in investigative effort. To minimize the intrusiveness of our active network measurements we implemented several procedures:

First, our research group incorporates an internal approval process before any measurement activities are carried out. This process is derived from University of Twente's process regarding ethical permissibility of research. This approval process, which involves multiple parties, allows us to reflect on the potential harm induced by network measurements and take preventive measures.

Second, we set up a website on the scanning machines which explains our measurement activity in detail.

[1] https://www.team-cymru.org/Services/Bogons/fullbogons-ipv6.txt

[2] https://www.iana.org/assignments/ipv6-address-space/ipv6-address-space.xml

TABLE I: IP address statistics for IXP and MWN.

| | IXP | | MWN | |
|---|---|---|---|---|
| Characteristic | Jul 28 - Aug 10, 2015 | Sep 03 - Sep 16, 2015 | Jul 28 - Aug 10, 2015 | Sep 03 - Sep 16, 2015 |
| Total IP address observations | 1,606,380,271 | 827,195,355 | 1,523,891,579 | 2,925,494,392 |
| Unique addresses | 70,288,801 (100%) | 80,121,373 (100%) | 4,797,664 (100%) | 5,901,149 (100%) |
| Removing Fullbogons | -2,656 (-0.00%) | -2,930 (-0.00%) | -45 (-0.00%) | -39 (-0.00%) |
| Removing IANA special | -1,315,539 (-1.87%) | -1,147,247 (-1.43%) | **-3,236,590 (-67.5%)** | **-4,610,134 (-78.1%)** |
| Removing pfx2as | -726 (-0.00%) | -844 (-0.00%) | -783 (-0.02%) | -863 (-0.01%) |
| Final | 68,969,997 (98.1%) | 78,970,545 (-98.6%) | 1,560,343 (32.5%) | 1,290,242 (21.9%) |
| Final, average per day: | 4,926,428 | 5,640,753 | 111,453 | 92,160 |

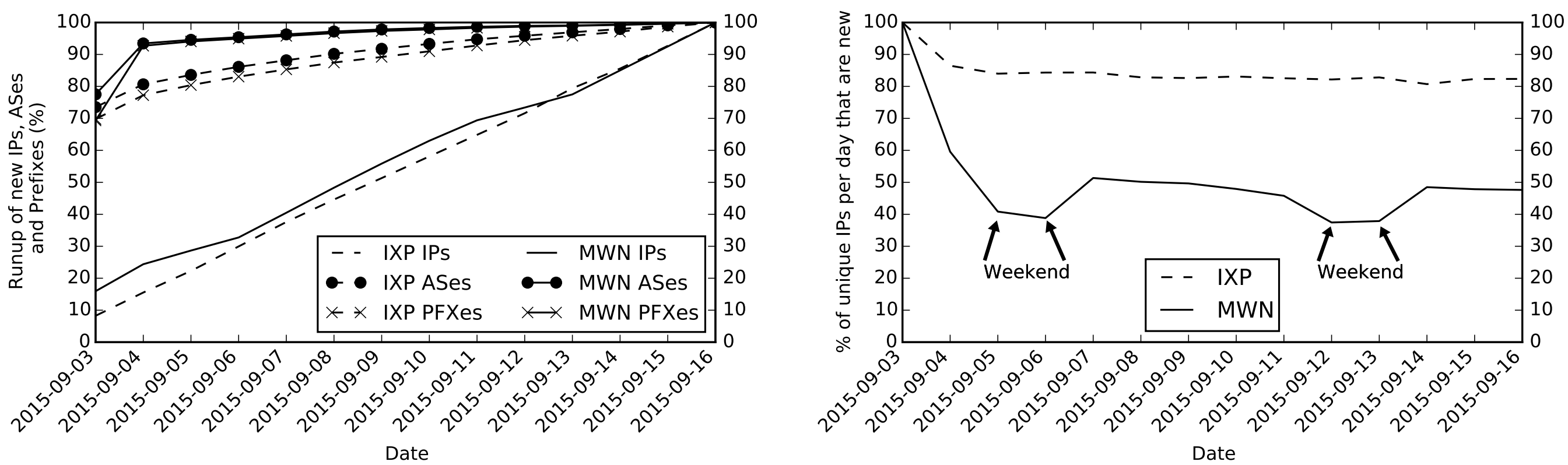

Fig. 1: IP, AS and prefix runup over two weeks.

Third, we maintain a blacklist of hosts and networks which will not be scanned in any of our measurements. Throughout the experiment, we received one e-mail out of curiosity and another one asking to be blacklisted. We complied with the blacklisting request and did not probe this network anymore. Partridge and Allman [20] propose to assess whether the collection of data can induce harm on individuals and whether the collected data reveals private information. By reducing the intrusiveness as discussed above we believe that no individual was harmed as a result of our active measurements. Furthermore, in order to protect private information we only release IPv6 hitlists generated from publicly available data on our website [11] and apply best practices [3].

## IV. Data Sources and Experiment

In this section we first describe the data sources that were used during the experiment. Then we give details about the capturing and preprocessing steps. Finally, we present evaluation results of the conducted experiments.

### A. Passive Sources

We obtained passive flow data from two collectors: A large European Internet Exchange Point (IXP) and the Internet uplink of the Munich Scientific Network (MWN) operated by the Leibniz Supercomputing Centre.

At the IXP, we obtained flow data at a sampling rate of 1:10,000 (systematic count-based sampling on packets). At the MWN, we received flow data for all packets. At each source we aggregated data into two sets, (a) from July 28 to August 10, 2015 and (b) from September 3 to September 16, 2015. The flows were filtered and then fed to a scanning engine at each source to repeatedly scan the observed IP addresses after specific time intervals (see Tables VI and VII). Detailed statistics are given in Table I: While the IXP observed a total of 146M unique addresses, the MWN network saw 2.7M unique addresses. The MWN network starts with a much higher count of IP address observations, but those quickly melt when made unique. Furthermore, more than two thirds of the observed addresses at MWN were filtered out as they were stemming from the IANA-special domain. This allows for two conclusions: First, uplink vantage points are less diversified in communication to unique IP addresses when compared to IXPs. Second, various tunneling techniques (subset of IANA reserved IPs) heavily distort the number of IP addresses observed at an uplink, but not at an IXP.

Figure 1 shows that the percentage of new IPs is close to linear for the IXP as well as the MWN. This hints at the usage of privacy extensions which will be evaluated in-depth later. After a couple of days we already have seen addresses from more than 90% of all observed ASes and prefixes. The dips in the right subfigure are weekend days and a result of the reduced presence and activity of researchers and students.

When looking at a port and protocol breakdown of the

TABLE II: Top 10 port-protocol combinations for IXP and MWN (based on count of flows).

| Rank | Jul 28, 2015 (IXP) | Jul 28 to Aug 10, 2015 (IXP) | Sep 03 to Sep 16, 2015 (IXP) | Jul 28, 2015 (MWN) | Jul 28 to Aug 10, 2015 (MWN) | Sep 03 to Sep 16, 2015 (MWN) |
|---|---|---|---|---|---|---|
| 1 | **tcp443 (31.65%)** | **tcp443 (31.42%)** | **tcp443 (34.31%)** | **tcp443 (20.05%)** | **tcp443 (20.91%)** | **tcp443 (19.84%)** |
| 2 | **tcp80 (13.10%)** | **tcp80 (13.88%)** | **tcp80 (10.64%)** | udp53 (13.59%) | udp53 (12.10%) | udp53 (12.21%) |
| 3 | udp53 (1.26%) | udp53 (1.17%) | udp53 (1.19%) | **tcp80 (9.49%)** | **tcp80 (9.40%)** | **tcp80 (10.74%)** |
| 4 | tcp119 (0.47%) | tcp119 (0.52%) | **udp443 (0.74%)** | icmp6 (3.03%) | icmp6 (3.42%) | icmp6 (1.68%) |
| 5 | **udp443 (0.45%)** | **udp443 (0.43%)** | icmp6 (0.38%) | **udp443 (1.50%)** | **udp443 (1.53%)** | **udp443 (1.37%)** |
| 6 | icmp6 (0.38%) | icmp6 (0.35%) | udp10000 (0.27%) | tcp5228 (0.94%) | tcp5228 (1.08%) | tcp5228 (0.92%) |
| 7 | udp10000 (0.36%) | udp10000 (0.32%) | tcp25 (0.25%) | tcp993 (0.74%) | tcp993 (0.75%) | tcp993 (0.64%) |
| 8 | tcp25 (0.24%) | tcp25 (0.17%) | tcp119 (0.23%) | udp123 (0.27%) | udp123 (0.36%) | udp123 (0.40%) |
| 9 | tcp22 (0.15%) | tcp993 (0.13%) | tcp993 (0.14%) | tcp25 (0.13%) | udp51413 (0.15%) | tcp53 (0.15%) |
| 10 | tcp993 (0.14%) | tcp22 (0.12%) | tcp22 (0.10%) | tcp143 (0.11%) | tcp25 (0.15%) | udp51413 (0.14%) |

TABLE III: AS[2] and prefix[2] statistics for IXP and MWN.

| Characteristic | IXP | MWN |
|---|---|---|
| File size | 5.3GB | 93MB |
| Input lines | 147M | 2.7M |
| Targets | **146,722,097** | **2,687,679** |
| ASes | 6,783 | 7,398 |
| Announced prefixes | 12,858 | 15,478 |
| AS coverage | **66.61%** | **72.65%** |
| ASes unique to source | 821 | 1,436 |
| Normalized ASes | 3,802 | 4,417 |
| Prefix coverage | **49.87%** | **60.04%** |
| Prefixes unique to source | 2,076 | 4,696 |
| Normalized prefixes | 7,467 | 10,087 |
| Combined AS coverage | 8,219 (80.71%) | |
| Combined prefix coverage | 25,781 (68.09%) | |
| ICMP response rate[1] ≈ | **13%** | **31%** |

1: Only referring to last two out of four weeks of observation at each source. Address considered responsive if a reply was received for any scan in the intervals shown in Tables VII and VI.
2: As found in the CAIDA prefix to AS mapping of Sep 07, 2015 [4].

observed flows, several characteristics stand out (c.f. Table II): First, the majority of flows stems from user-generated traffic (tcp80, tcp443), with tcp443 showing clear dominance over tcp80 and further increasing over time. Second, udp443, likely traffic caused by Google's QUIC protocol [25], sees a relevant and increasing share during the observation period. As a difference between the sources, a significantly higher share of DNS flows at MWN stands out. This is to be expected, as all up- and downstream recursive DNS is processed through this link, whereas most DNS requests at an IXP are likely routed through the peers' regular Internet uplinks.

Table III compares the covered ASes and prefixes of the two passive sources. Even though we see magnitudes more IPs at the IXP, the coverage of ASes as well as prefixes is larger for the MWN source. Moreover, the response rate to ICMP echo requests is much higher at the MWN than the IXP.

### B. Active Sources

As an active source we used the Alexa Top 1 Million list [2] where we queried the domain names for AAAA DNS records. Another source is a complete reverse lookup of the entire IPv4 address space provided by Rapid7 [24]. We extracted the hostnames from this list and recursively queried those for AAAA records. Additionally Rapid7 offers the DNS ANY Record file [23]. It contains replies from large-scale DNS ANY scans, which we subsequently filtered for IPv6 addresses. Another data set are zone files for several top level domains [22] which we resolved for their AAAA records. All DNS queries were sent from a server at our chair. We also used the CAIDA DNS names data set [5] containing IPv6 addresses obtained by traceroutes to random targets. All sources except the CAIDA DNS names data set are compared in Table IV. All active sources hitlists are available on our website [11].

Regarding ASes and prefixes we first look at the coverage, i.e. the percentage of ASes and prefixes covered in each source compared to all announced ASes and prefixes. Furthermore, we identify ASes and prefixes which are uniquely covered in only one source. In addition we provide another metric, the *normalized* ASes and prefixes: We assign a weight to each AS or prefix, which is defined as the inverse of the number of sources it is contained in. We then sum up the weights for all ASes and prefixes per source. Finally, we also provide response rates for the various scan types and give a combined coverage of ASes and prefixes. One takeaway from the comparison is that the rDNS source provides little added value given the effort to resolve and then parse more than one billion host names to gain just 30 unique ASes. Furthermore, the response rate of the found hosts is generally high. Surprisingly ICMPv6 gives the most replies, even for the Alexa list which should exclusively contain web servers responsive on port tcp80.

### C. Traceroutes to Target Lists

We also set out to answer how many more addresses can be discovered by conducting traceroute measurements to known existent addresses. Using scamper [18], we conducted path measurements from hosts in Munich, Singapore and Dallas. From every source, we probed all addresses learned from active sources. These yielded more addresses from intermediate routers, depicted in Table V. This table also provides a comparison of the new addresses learned from our traceroutes

TABLE IV: Analysis of active sources.

| | Alexa Top 1M | | rDNS | | DNS Any | | Zone Files | |
|---|---|---|---|---|---|---|---|---|
| File size | 22MB | | 56GB | | 69GB | | 2.6GB | |
| Input lines | 1M | (100%) | **1.2G** | (100%) | **1.4G** | (100%) | 151M | (100%) |
| Raw addresses | 90,671 | (9.07%) | 1,023,950 | (0.08%) | 9,768,810 | (0.68%) | 4,762,297 | (3.14%) |
| Targets | 43,822 | (4.38%) | 462,185 | (0.04%) | 1,440,987 | (0.10%) | 424,748 | (0.28%) |
| ASes | 1,424 | | 4,795 | | 5,708 | | 2,371 | |
| Prefixes | 1,695 | | 6,749 | | 8,506 | | 2,995 | |
| AS coverage[1] | 14.0% | | 47.1% | | 56.1% | | 23.3% | |
| ASes unique to source | 1 | | **30** | | **685** | | 5 | |
| Normalized ASes | 401.3 | | 1,919.5 | | 2,694.8 | | 734.3 | |
| Prefix coverage[1] | 6.57% | | 26.2% | | 33.0% | | 11.62% | |
| Prefixes unique to source | 7 | | 65 | | **1,379** | | 11 | |
| Normalized prefixes | 490.7 | | 2,816.7 | | 4,338.8 | | 955.8 | |
| ICMPv6 response rate | 41,759 | **(95.3%)** | 317,773 | **(68.8%)** | **1,046,562** | **(72.6%)** | 385,023 | **(90.6%)** |
| tcp80 response rate | 41,279 | (94.2%) | 131,403 | (28.4%) | 744,100 | (51.6%) | 375,052 | (88.3%) |
| tcp443 response rate | 33,225 | (75.8%) | 98,174 | (21.2%) | 400,182 | (27.8%) | 249,112 | (58.6%) |
| Combined AS coverage | | | | **7,331 (71.9%)** | | | | |
| Combined prefix coverage | | | | **12,854 (49.8%)** | | | | |

1: Compared to the CAIDA prefix to AS mapping of Sep 07, 2015 [4].

TABLE V: Analysis of new IPs found through traceroute measurements.

| | Acquired from traceroutes on active sources | | | | | | Self-standing source | Merged |
|---|---|---|---|---|---|---|---|---|
| | Alexa Top 1M | rDNS | DNS Any | Zone Files | CAIDA DNS-Names | Merged | CAIDA DNS-Names | Merged |
| Raw IPs | 155,046 | 9,681,039 | 28,260,818 | 14,249,257 | 315,123 | n/a | 1,236,960 | n/a |
| Filtered IPs[1] | 50,479 | 366,183 | 1,161,900 | 416,843 | 90,848 | 1,259,283 | 102,580 | 183,067 |
| New IPs only | 8,742 | 49,549 | 91,445 | 26,870 | 11,714 | 109,554 | 102,580[2] | 183,067 |
| ASes | 1,216 | 3,354 | 3,928 | 2,012 | 1,014 | **4,170** | **5,488** | **6,287** |
| Prefixes | 1,439 | 4,178 | 5,007 | 2,498 | 1,176 | **5,367** | **9,269** | **10,466** |
| AS coverage[3] | 11.9 % | 32.9 % | 38.6 % | 19.7 % | 10.0 % | 41.0 % | 53.9 % | 61.7 % |
| Prefix coverage[3] | 5.6 % | 16.2 % | 19.4 % | 9.7 % | 4.6 % | 20.8 % | 36.0 % | 40.6 % |

1: Following the filtering cascade described in section III. 2: Assuming CAIDA data set includes routers only and not the original target IPs.
3: Compared to the CAIDA prefix to AS mapping of Sep 07, 2015 [4].

to those learned from CAIDA's traceroutes. One quickly sees that CAIDA's approach of probing random IP addresses in all announced prefixes yields a great AS and prefix coverage that exceeds even those from all our measurements combined. However, tracerouting adds another 80,487 (78%) router IP addresses and 7.8% on AS coverage when compared to only using the CAIDA DNS-Name data set.

### D. Source Diversity

We understand that measurement results heavily depend on the input data. Therefore we survey the diversity of our active, passive and traceroute sources to ascertain their diversity.

The passive sources' flow data are obtained at specific locations: The MWN source is located in Munich, Germany. However, it includes heterogeneous traffic from multiple universities, research institutes and end-users in the form of student dormitories. The IXP source on the other side stems from a large European IXP. This IXP's customers include stub ASes, transit ASes as well as ISP ASes offering end customer access. As a result our passive sources provide a heterogeneous mix of European-centered traffic.

Our active sources are the Alexa Top 1 Million list, rDNS, ANY DNS, CAIDA DNS names, and DNS zone files. Due to their inherent geographic distribution the first four sources can be seen as substantially diverse. The addresses they contain are not limited to any one specific network or parts of the Internet, but represent the most common addresses. The DNS zone files, however, are limited to the zones contained in this source. This might introduce a slight bias towards addresses in North America as major European country top level domains (TLDs) such as `.de`, `.co.uk`, `.fr`, and `.it` are missing. Since the organizations administrating these TLDs do not release their zone files we have to take note of this slight bias.

From tracerouting we receive additional router addresses. These addresses are located in 4,170 ASes and 103 countries, which we consider a good coverage for a router only data set.

We conclude that the mix of obtained addresses is heterogeneous and representative for IPv6 traffic in the Internet.

## V. Evaluation

In this chapter we evaluate our sources along different criteria: First, we assess the response rate of IPv6 addresses over time. Second, we survey MAC addresses encoded in interface IDs and the usage of privacy extensions. Then we provide insights into the number of IP addresses as a metric for IPv6 hitlists. Finally, we give concrete advise on which source to use for hitlist generation depending on the scan type.

### A. Response Rates over Time

At the passive sources, we performed active scans towards each observed IP address both using ICMP and the protocol it

TABLE VI: MWN response rates over time after first address observation.

| Scan Type | # Targets | 1 minute | 10 minutes | 1 hour | 12 hours | 1 day | 3 days | 7 days |
|---|---|---|---|---|---|---|---|---|
| icmp6 | 828,142 | 358,771 (43.32%) | 349,030 (42.14%) | 331,795 (40.06%) | 296,657 (35.82%) | 288,637 (34.85%) | 282,622 (34.12%) | 276,864 (33.43%) |
| tcp80 | 82,015 | 78,280 (**95.44%**) | 78,268 (95.43%) | 78,307 (95.47%) | 78,187 (95.33%) | 78,024 (95.13%) | 78,027 (95.13%) | 77,456 (**94.44%**) |
| tcp443 | 82,015 | 59,151 (72.12%) | 59,126 (72.09%) | 59,121 (72.08%) | 59,071 (72.02%) | 58,841 (71.74%) | 58,894 (71.80%) | 58,256 (71.03%) |
| udp443 | 5,292 | 3,211 (60.67%) | 3,218 (60.80%) | 3,228 (60.99%) | 3,218 (60.80%) | 3,210 (60.65%) | 3,226 (60.95%) | 3,161 (59.73%) |
| udp49001 | 7,314 | 3,748 (**51.24%**) | 3,462 (47.33%) | 2,698 (36.88%) | 860 (11.75%) | 778 (10.63%) | 602 (8.23%) | 547 (**7.47%**) |
| udp51413 | 12,875 | 5,488 (42.62%) | 5,359 (41.62%) | 5,035 (39.10%) | 4,356 (33.83%) | 4,242 (32.94%) | 4,018 (31.20%) | 3,852 (29.91%) |

TABLE VII: IXP response rates after first address observation.

| Scan Type | # Targets | 1 minute | 1 hour | 1 day | 7 days |
|---|---|---|---|---|---|
| icmp6 | 66,079,853 | 8,780,586 (**13.28%**) | 2,691,658 (4.07%) | 905,833 (1.37%) | 622,412 (**.94%**) |
| tcp80 | 392,913 | 276,964 (70.48%) | 263,088 (66.95%) | 243,693 (62.02%) | 236,122 (60.09%) |
| udp443 | 2,839 | 2,317 (81.61%) | 1,876 (66.07%) | 1,806 (63.61%) | 1,584 (55.79%) |
| udp49001 | 25,145 | 14,500 (**57.66%**) | 7,893 (31.38%) | 1,216 (4.83%) | 559 (**2.22%**) |
| udp51413 | 32,732 | 4,125 (12.60%) | 2,979 (9.10%) | 1,751 (5.34%) | 1,392 (4.25%) |

TABLE VIII: In-protocol vs. ICMPv6 response rates.

| | IXP | | MWN | |
|---|---|---|---|---|
| Scan Type | Replies[1] | ICMP unresponsive | Replies[1] | ICMP unresponsive |
| icmp6 | 8,710,139 | n/a | 257,563 | n/a |
| tcp80 | 180,577 | 2,387 (**1.32%**) | 54,010 | 937 (**1.73%**) |
| tcp443 | n/a | n/a | 40,439 | 1,108 (2.74%) |
| udp443 | 2,192 | 81 (3.70%) | 2,945 | 0 (0.00%) |
| udp49001 | 14,343 | 10,361 (**72.2%**) | 3,709 | 2,391 (**64.5%**) |
| udp51413 | 3,514 | 274 (7.80%) | 4,580 | 83 (1.81%) |

1: Unique IPs which responded to in-protocol measurement.

was seen on. Those were repeated at certain intervals. At the larger IXP source, we spared some time and protocol slots to fit our bandwidth limits.

The Tables VI and VII and Figure 3 show the response rates over time for MWN and IXP. The IPs learned at MWN show a more time-stable behavior. Server-type ports (tcp80, tcp443, udp443) show a very high and stable response rate. BitTorrent (udp49001) shows a strong decrease after an initially high response rate, while Mainline DHT (udp51413) has a more stable response rate. Passive sources might include spoofed addresses resulting in a lower response rate. Furthermore, ICMPv6 is quite stable at MWN, whereas with a lower response rate at the IXP. One possible explanation is rate limiting of ICMPv6 packets at routers in target networks due to the two magnitudes higher packet volume at the IXP compared to MWN. In addition to observed ports all hosts were queried for ICMPv6.

Table VIII compares response rates between in-protocol scans (i.e., an observed host is scanned on the port and protocol of the observation) and a generic `ICMPv6 Echo Request` scan. The results very clearly show that generally almost all hosts that reply in-protocol will also reply to ICMP echo requests. One remarkable exception is udp49001, where over 60% of hosts did not reply to echo requests. This is likely due to udp49001 being a BitTorrent port and the associated hosts may be behind home routers, blocking ICMP.

### *B. Interface Identifier Analysis*

To better understand the acquired addresses we analyze the interface identifier (IID) of these addresses, i.e. the last 64 bit.

TABLE IX: Top 5 vendors for EUI-64 IPs.

| | IXP | | Scamper | |
|---|---|---|---|---|
| Position | Vendor | Percentage | Vendor | Percentage |
| 1 | Samsung | 30.7% | Arcadyan | 28.4% |
| 2 | Apple | 11.6% | Huawei | 24.4% |
| 3 | Sony | 5.8% | AVM | 16.0% |
| 4 | Murata | 5.1% | Sercomm | 10.5% |
| 5 | Huawei | 5.1% | Cisco | 4.4% |

**Modified EUI-64 ID.** IPv6 has a mechanism to automatically assign addresses without a DHCPv6 server. This *Stateless Address Autoconfiguration* mechanism usually takes an interface's MAC address and modifies it by inserting `ff:fe` in the middle and flipping the 6th bit (RFC 4291). This modified EUI-64 ID is then appended to the announced prefix.

In total we found about 2.5M EUI-64 IPs. IPs from the active sources (8.6%) and Scamper (7.0%) are more likely to be EUI-64 IPs than MWN (3.9%), IXP (1.6%) and CAIDA (1.4%), likely due to multiple counts of clients using privacy extensions at passive sources.

We aggregated the EUI-64 IDs by MAC vendors. Table IX shows the top 5 vendors for the IXP and Scamper sources. Not surprisingly, the IXP's EUI-64 addresses are mainly from user devices while the top 5 Scamper EUI-64 vendors are networking equipment manufacturers. MWN and IXP as well as Scamper and CAIDA show similar distributions.

**Privacy Extensions.** To avoid unique traceability through MAC addresses encoded within the IID, RFC 4941 presents *Privacy Extensions*. These reduce traceability by randomizing the IID. Since the 6th bit (leftmost bit is 0) of the IID is always set to 0 indicating local scope, 63 uniformly distributed bits remain for the IID. By applying the *central limit theorem*, the sum of these single bit distributions approximates the normal distribution $\mathcal{N}(31.5, 15.75)$. Figure 2 shows the Hamming weight distribution for the IXP and Scamper sources.

The Hamming weight distribution at the IXP approximates the aforementioned normal distribution. This clearly indicates that the vast majority of the IXP's addresses contain a random IID. A similar phenomenon was observed for the MWN source. Addresses obtained from Scamper on the other hand differ drastically: Two thirds of IIDs have less than six bits set to one, with more than 40% only having one bit set. This hints at a large number of statically assigned addresses which seems reasonable for a source primarily consisting of routers.

**Prefix Agility.** The movement of IIDs between prefixes is called prefix agility. To measure prefix agility, we evaluate the number of IIDs which were seen in more than one /64 prefix. IIDs from passive sources are relatively agile with ≈50% appearing in 2 or more prefixes. These findings are different to Plonka and Berger's [21] who find ≈12% of modified EUI-64 IIDs being agile. For active sources and Scamper, we found only about 2% of IIDs to be agile. CAIDA showed a surprisingly high agility of 14%, likely due to routers multi-homing in multiple /64s.

We acknowledge that IID analysis could be influenced by chance or willful action of a network administrator. However, we consider this effect minor for large-scale analysis.

### *C. The Value of IPs as a Metric*

Some insights throughout this paper show that a simple count of IP addresses is not valuable: First, privacy extensions dominate the observations at both passive sources and keep the IP count growing almost linearly, while the number of Autonomous Systems and prefixes found quickly saturates. Second, few IPs form a stable core and are frequented by many clients: the flow statistics show that 40-45% of flows are observed on tcp443 or tcp80. As these flows typically have a high (≥1024) port number on the other end, one can derive that almost all traffic (80-90%) is to or from a set of HTTP and HTTPS servers. However, Tables VI, VII and VIII show that only ≈12% (MWN), respectively ≈0.6% (IXP) of targets are associated to server ports, which will likely be more responsive throughout time. Third, this stable core covers a significant part of ASes and prefixes: Table X shows the AS coverage by servers only. We find that servers cover ≈30-50% of prefixes and ≈50-70% of ASes. This is due to an AS announcing several prefixes. Covering one of these prefixes already results in counting the AS.

### *D. Specific Approaches Depending on Scan Type*

In this paper we present many different sources to create a hitlist for IPv6 measurements, aggregated in Table XI. However, the value and usefulness of each of these sources depends on the research question to be answered and consequently the type of scan to be carried out. Moreover, not all of these sources might be available to every researcher. In addition, the effort in terms of data storage, processing power and network bandwidth should not be underestimated. Therefore we dedicate this section to recommend the most efficient combination of sources tailored specifically to the type of scan in question.

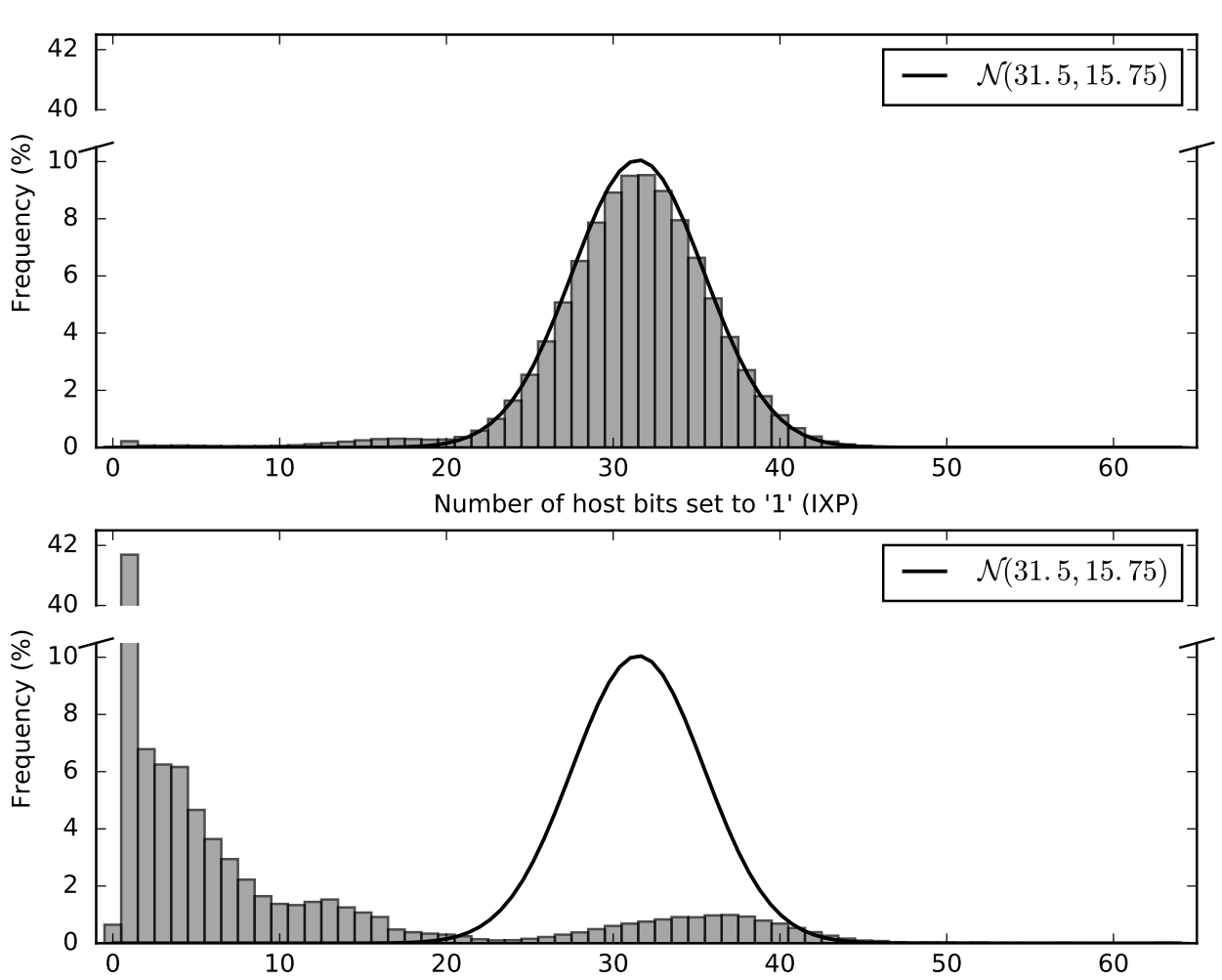

Fig. 2: Hamming weight distribution of interface ID.

TABLE X: AS and prefix coverage by servers.

| Scan Type | IXP Servers | IXP Total | MWN Servers | MWN Total |
|---|---|---|---|---|
| ASes | 4,452 (65%) | 6,783 | 3,909 (53%) | 7,398 |
| Prefixes | 6,736 (52%) | 12,858 | 5514 (35%) | 15478 |

TABLE XI: Statistics of active, passive, and traceroute sources.

| Characteristic | Active sources | Passive sources | Traceroutes | CAIDA [5] |
|---|---|---|---|---|
| File size | 75MB | 5.4GB | 2.4MB | 40MB |
| Unique input lines | 2.7M | 149M | 1.3M | 618k |
| Unique targets | 2,699,573 | 148,631,234 | 109,554 | 102,580 |
| Unique ASes | 5,750 | 8,219 | 4,170 | 5,488 |
| Unique announced prefixes | 8,602 | 17,554 | 5,367 | 9,269 |
| AS coverage | 56.46% | **80.71%** | 41.00% | **53.90%** |
| ASes unique to source | 128 | 1,276 | 14 | 147 |
| Normalized ASes | 1,918.33 | 3,684.67 | 1,158.83 | 1,873.17 |
| Prefix coverage | 33.37% | **68.09%** | 20.76% | **36.00%** |
| Prefixes unique to source | 346 | 5,798 | 53 | 514 |
| Normalized prefixes | 3,199.25 | 10,302.58 | 1,569.92 | 3,681.25 |
| ICMPv6 response rate | 75.5% | 13.3% | n/a | 42.0% |
| Combined unique IPs | **149,619,624** | | | |
| Combined AS coverage | **8,531 (83.77%)** | | | |
| Combined prefix coverage | **18,502 (71.77%)** | | | |

**Internet structure:** Evaluating the Internet structure aims at finding as many routers and transit links as possible. Therefore, it is of essence to maximize the count of ASes and announced prefixes in the hitlist (in contrast to maximizing IP count). Table XI shows that a combination of passive sources and the CAIDA DNS dataset [5] yields very high AS and prefix coverage at low effort. Prefixes missing from that combination could be probed using guessed IIDs (e.g. `::1`).

**Assessing security posture:** Empirically assessing the Internet's security posture aims at scanning as many responsive hosts as possible, although frequently only servers are of interest. We recommend to start with active sources, quickly providing 2.7M unique targets which are likely servers and 75% responsive. To further extend coverage, passive sources could be leveraged. We conducted a more specific evaluation for web servers (tcp80, tcp443 and udp443) in Table XII. It shows that when looking for web server IP addresses, active sources provide the bulk mass of unique addresses. However, passive sources, if available, are still a reasonable addition in terms of addresses, prefixes and ASes covered.

**Internet routers:** When aiming to scan Internet routers, we advise to use CAIDA's IPv6 DNS dataset as the first source. As seen in Table V it provides very good coverage with very low effort. When aiming for maximum coverage, traceroutes to other active sources will incrementally add more IP addresses.

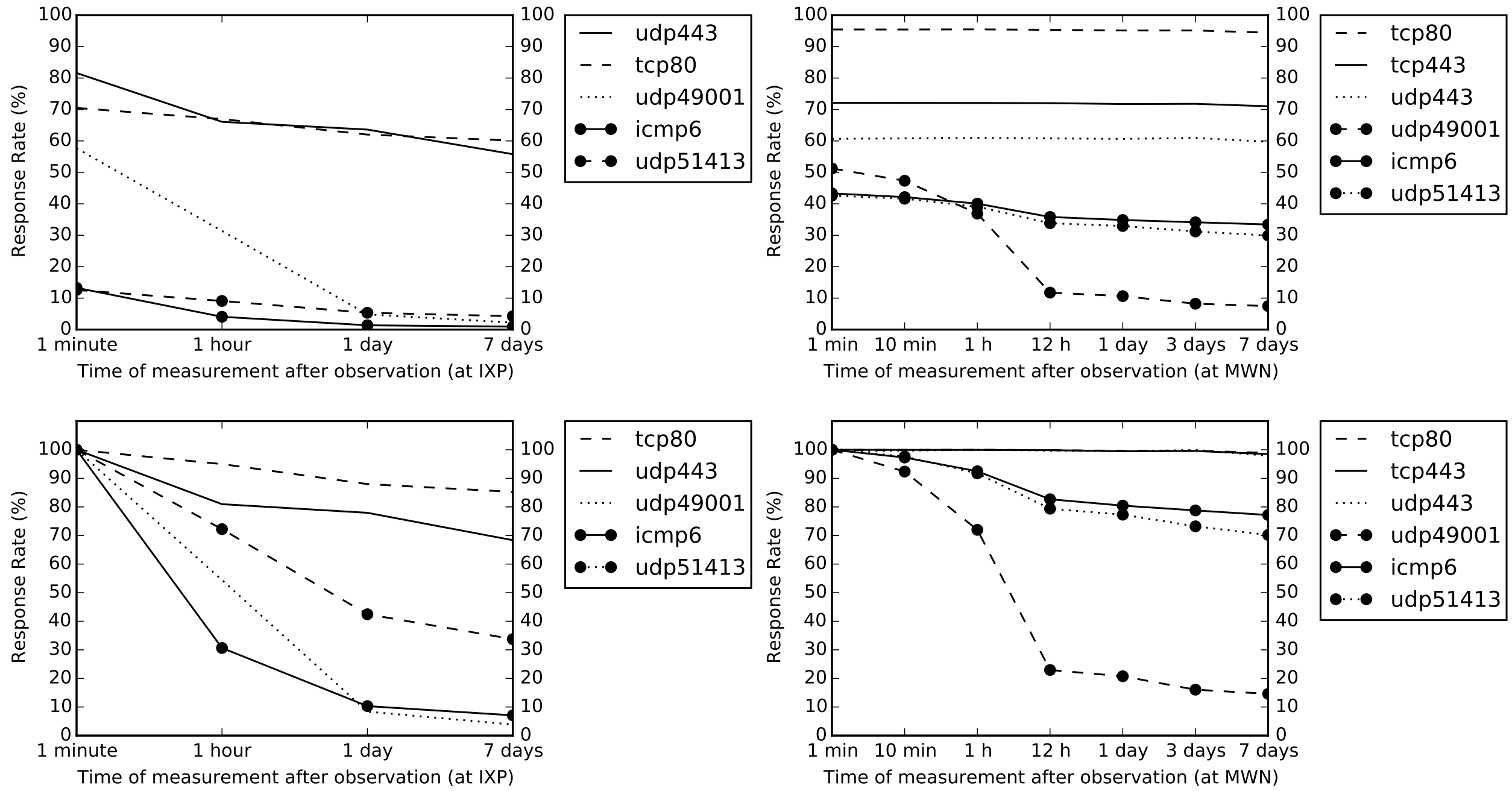

Fig. 3: Response rate related to observed addresses (top row) and responsive addresses (bottom row).
Note: icmp6 rate related to all addresses, port/protocol rate related to addresses observed on port/protocol.

TABLE XII: Web Server statistics.

| | IXP[1] | MWN[1] | Alexa | Zone Files |
|---|---|---|---|---|
| Unique filtered addresses | 256,891 | 56,846 | 43,822 | 752,585 |
| IPs unique to source | 86,996 | 5,573 | 4,180 | 416,477 |
| ASes unique to source | 325 | 59 | 101 | 784 |
| Prefixes unique to source | 650 | 111 | 144 | 1,100 |

1: Addresses observed on ports tcp80, tcp443 or udp443.

**Clients:** When analyzing clients, a passive tap is a good way to gather active addresses (see Tables VII and VI). However, these quickly vanish, so almost immediate scans are advisable.
**Active prefixes:** Passive sources are key to identifying active prefixes and their subprefixes.

## VI. Conclusion and Future Work

We leveraged both passive and active measurements to gather 150M IPv6 addresses covering 84% of ASes and 72% of prefixes. We actively probed these addresses to evaluate the response rate. We found very diverse characteristics of sources with regards to coverage and efficiency. We derived specific recommendations which source to use for different scan types. Additionally, we argue that the sole number of IPv6 addresses on a hitlist is not very relevant and can be vastly misleading. Therefore, we suggest to build a hit list focused on stable IP addresses covering a diverse set of ASes and prefixes.

**Future Work.** To further tailor the hit list to a specific scan type a classification of hosts (e.g. server, router, client) could help. Moreover, one could try to derive patterns from stable IP addresses to predict responsive addresses in other subnets.

**Acknowledgments.** We thank the IXP and the Leibniz Supercomputing Centre for providing the flow data used in the experiments. This work was supported by the German Federal Ministry of Education and Research, project Peeroskop, grant 01BY1203C, and project SURF, grant 16KIS0145, and by the European Commission, project SafeCloud, grant 653884.